# Kin: the optimal variable for CP-violation

S. Henrot-Versillé[1], F. Le Diberder[2]

[1] Laboratoire de l'Accélérateur Linéaire, Orsay, France
[2] Laboratoire de Physique Nucléaire des Hautes Energies, Paris, France

**Abstract**

We advocate the use of an optimal variable to demontrate CP violation. This variable, named the Kin variable, provides a means to visualize graphically CP violation, to pin point the most statistically powerful events, and to combine several channels and experiments. For the $B^0 \to J/\Psi K_S^0$ channel, its use leads to a straightforward measurement of $\sin(2\beta)$.

The demonstration of CP violation with $B$ decays, followed by the detailed studies of its pattern, is a crucial goal for the coming years. The two asymmetric B-factories, PEP-II and KEKB, started to deliver data to the BaBar and Belle detectors in 1999: both experiments are expected to provide their first results before the end of the year 2000. Observation of CP violation should be achieved, notably with the golden channel $B^0 \to J/\Psi K_S^0$, by the measurement of the CP-violating parameter $\sin(2\beta)$.

The initial analyses will deal with low statistic samples of events, and the question arises of how to exhibit the presence of CP violation in the data. Whereas the measurement of $\sin(2\beta)$ is simply obtained using the likelihood method, this mesurement, by itself, does not provide a clear view of where CP violation resides in the data: it lacks a straightforward graphical display of CP violation.

We introduce below a variable, named Kin† and denoted $\mathcal{K}$, which is optimal for CP violation. The variable is claimed to be optimal because the $\sin(2\beta)$ measurement relies only on the value it takes per event. The distribution of the Kin variable provides a means to demonstrate CP violation and to perform a straightforward measurement of $\sin(2\beta)$, with an accuracy almost identical to the one achieved by the likelihood method. Similar approaches leading to the definition of optimal variables have already been used for other types of analyses; for the measurement of the $\tau$ polarization[1], and for search of CP violation in $\tau$ decays [2].

After a brief introduction to the notations, we describe the Kin variable, and show how to demonstrate CP violation and measure $\sin(2\beta)$ with its use.

† The name arises from the japanese: the Kin (金) variable means the gold variable in Japanese.

## 1. Definition of the Kin variable

For the sake of simplicity, and since these channels are the ones expected to lead to the first observation of CP violation in $B$ decays, we consider an experiment carried on in a B-factory analyzing channels related to $\sin(2\beta)$ through time dependent decay rates of the form:

$$\rho(t) = \frac{\Gamma}{2} e^{-\Gamma|t|} \left(1 + s \, \sin(2\beta)\sin(\Delta m_d t)\right) \quad , \quad (1)$$

where $t$ is the $B^0 \to J/\Psi K_S^0$ decay time‡, and $s = 1$ (resp. $s = -1$) if the neutral $B$ at $t = 0$ is a $B^0$ (resp. a $\bar{B}^0$), as determined by tagging the heavy quark content of the companion $B$ meson. One denotes $\overline{\rho}(t)$ the time dependent decay rates of the CP conjugate decay:

$$\overline{\rho}(t) = \frac{\Gamma}{2} e^{-\Gamma|t|} \left(1 - s \, \sin(2\beta)\sin(\Delta m_d t)\right) \quad . \quad (2)$$

Since CP violation is established by the observation of $\rho(t) \neq \overline{\rho}(t)$, and since $\sin(2\beta)$ enters linearly in Eqs. (1)-(2) we introduce the Kin variable $\mathcal{K}$ defined by:

$$\sin(2\beta)\mathcal{K} = \frac{\rho - \overline{\rho}}{\rho + \overline{\rho}} \quad . \quad (3)$$

In the absence of backgrounds and of detector effects, $\mathcal{K}$ only depends on $t$ and on $s$, and takes its values in the interval $[-1, +1]$:

$$\mathcal{K} = s \, \sin(\Delta m_d t) \quad . \quad (4)$$

When backgrounds and detector effects are taken into account, the expected time distribution of events, still denoted $\rho(t)$, differs from Eq. (1), but the two above key points remain true: CP violation is equivalent to $\rho(t) \neq \overline{\rho}(t)$, and $\sin(2\beta)$ enters linearly in the expression of $\rho(t)$. The Kin definition of Eq. (3) still applies, but it is no longer identical to Eq. (4), and the interval where $\mathcal{K}$ takes its values is reduced (cf. Eqs(11-12) of section 3).

‡ In BaBar and Belle experiments, $t$ is obtained as the difference between the decay times of the two neutral $B$'s.

## 2. Properties of the Kin variable

The definition (Eq. (3)) of the $\mathcal{K}$ variable, implies the following properties:

• Because of the trivial identity

$$\rho = \frac{1}{2}(\rho + \bar{\rho})(1 + \sin(2\beta)\mathcal{K}) \quad , \quad (5)$$

and since, by construction, the sum $\rho + \bar{\rho}$ is CP invariant, a likelihood analysis of a sample of $N$ events only depends on the Kin value per event $\mathcal{K}_i$:

$$\mathcal{L}(\sin(2\beta)) = \sum_{i=1}^{N} \ln(\rho(t_i)) \quad (6)$$

$$= \sum_{i=1}^{N} \ln(1 + \sin(2\beta)\mathcal{K}_i) + cst \quad . \quad (7)$$

Hence, the information on CP violation is fully contained in the Kin values.

• The Kin distribution, denoted $\Psi(\mathcal{K})$, takes the form:

$$\Psi(\mathcal{K}) = \Psi_0(\mathcal{K})[1 + \sin(2\beta)\mathcal{K}] \quad (8)$$

where $\Psi_0$ is an even function of $\mathcal{K}$. Hence, the observation of an asymmetry in the $\mathcal{K}$ distribution establishes (graphically) the occurence of CP violation in the data. As far as the detector and the analysis do not simulate CP violating effects, the net impact of experimental limitations is to modify $\Psi_0(\mathcal{K})$ (cf. Eqs.(11-12) of section 3), however keeping its key property: it remains an even function of $\mathcal{K}$.

• Because $\Psi_0(\mathcal{K})$ is an even function of $\mathcal{K}$, the value of $\sin(2\beta)$ can be inferred from the first two moments of the distribution:

$$\sin(2\beta) = \frac{\langle \mathcal{K} \rangle}{\langle \mathcal{K}^2 \rangle} \quad . \quad (9)$$

Hence, the measurement is straightforward: no likelihood maximization is needed.

• The statistical accuracy of the above $\sin(2\beta)$ measurement is given by:

$$\sigma[\sin(2\beta)] = \frac{1}{\sqrt{N\langle \mathcal{K}^2 \rangle}}\sqrt{1 - \sin^2(2\beta)\frac{\langle \mathcal{K}^4 \rangle}{\langle \mathcal{K}^2 \rangle}} \quad . \quad (10)$$

This expression differs from the one of a likelihood analysis by higher order moments ($\langle \mathcal{K}^{\geq 6} \rangle$). Hence, for all practical purpose, the two uncertainties are numerically identical (cf. section 3).

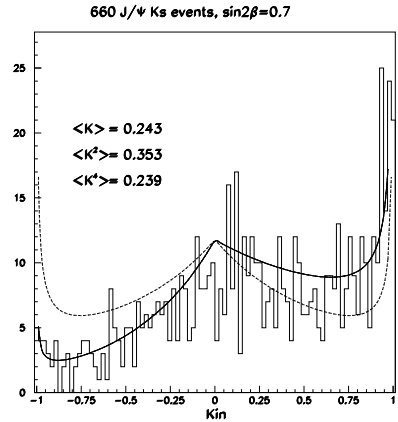

**Figure 1.** *The histogram corresponds to the Kin distribution of 660 $B^0 \to J/\Psi K_S^0$ events simulated with $\sin(2\beta) = 0.7$. Superimposed in dashed lines is $\Psi_0(\mathcal{K})$ (no CP violation in the data) while the solid line represents $\Psi(\mathcal{K})$ (cf. Eq. (8)). A measurement of $\sin(2\beta)$ is provided by the ratio $\langle \mathcal{K} \rangle / \langle \mathcal{K}^2 \rangle$ (see text).*

• Since the Kin variable is defined on an event-by-event basis, it provides a means to isolate events the most sensitive to CP violation. One may rank events according to their contribution to the measurement of $\sin(2\beta)$, that is to say (cf. Eq. (10)) according to their $\mathcal{K}^2$ value. Events with a Kin value close to zero do not carry information on CP violation. Hence, the most potent events are located in the wings of the $\mathcal{K}$ distribution: those are the ones to be studied the most carefully, the ones from which systematical biases may arise.

## 3. Illustrations

The distribution of the Kin variable is shown on figure 1 for the ideal case where Eq. (1) holds. The dotted curve is $\Psi_0(\mathcal{K})$ (i.e no CP violation, $\sin(2\beta) = 0$) and the solid curve is $\Psi(\mathcal{K})$ (with $\sin(2\beta) = 0.7$). The histogram results from a Monte Carlo simulation of a sample of 660 $J/\Psi(l^+l^-)K_S^0(\pi^+\pi^-)$ events†. From the moments of the distribution one gets: $\sin(2\beta) = 0.69 \pm 0.05$.

In the presence of background and detector effects, the time distribution of events is no longer given by Eq. (1), but becomes:

$$\rho = \mathrm{b}(y)\rho_{\exp}(t) + \overline{\mathrm{b}}(y)\bar{\rho}_{\exp}(t) + \mathrm{q}(y)\rho_{\mathrm{bkg}}(t) \quad (11)$$

$$\bar{\rho} = \overline{\mathrm{b}}(y)\rho_{\exp}(t) + \mathrm{b}(y)\bar{\rho}_{\exp}(t) + \mathrm{q}(y)\rho_{\mathrm{bkg}}(t) \quad (12)$$

where:

† This corresponds roughly to an integrated luminosity of 30 fb$^{-1}$ at a B-factory [3].

- b, $\overline{\text{b}}$, and q are the relative probabilities that the event corresponds at $t = 0$ to a $B^0$, a $\bar{B}^0$, or is a background event. By construction they satisfy b + $\overline{\text{b}}$ + q = 1. The notation y stands for the set of discriminating variables upon which are based the computation of the above quantities.

- $\rho_{\text{exp}}$ and $\bar{\rho}_{\text{exp}}$ are obtained through a convolution of the expressions of Eq. (1) and Eq. (2) by the detector response, while $\rho_{\text{bkg}}$ is the time distribution of backgrounds, taking into account detector effects.

We consider below three experimental limitations. Figure 2 shows a set of expected Kin distributions for 400 tagged $B^0 \to J/\Psi K_S^0$ events (i.e. assuming that a 60% tagging efficiency must be applied to the 660 events of figure 1). The curve is the previous $\Psi(\mathcal{K})$ function, scaled to 400 events. In the simulation, the vertex resolution is taken to be $\sigma_z = 120 \mu$m, the background over signal ratio is $B_S = 0.06$, and the mistag rate is taken to be the constant $w_{\text{tag}} = 0.15$. The three effects are taken into account one after the other, in the above order. One observes that the Kin distribution shrinks toward the origin with the result that $\langle \mathcal{K}^2 \rangle$ decreases, and that $\sigma[\sin(2\beta)]$ increases accordingly (cf. Eq. (10)). The most significant effect comes from the non-zero mistag rate.

by experiment) analyses, a global analysis can be carried out straightforwardly that way. This is illustrated on figure 3 where the 13 channels listed in table 5.38 of Ref. [3] have been merged together (the limitations due to background, tagging and vertexing are also taken into account here). A transversity analysis is used for vector-vector final states. This is taken into account in a transparent way in the computation of the Kin variable, which then also depends on the kinematics of the decays. Similarly, we use here a single value for the mistag rate, but one can define it per tagging category (lepton tagging, kaon tagging, etc.) or, more generally, on an event to event basis using a set of discriminating variables: the Kin values will then depend on the tagging category on the discriminating variables.

The curve of figure 3 represents the $\Psi_0(\mathcal{K})$ function. The departure of the event distribution from this curve is the signal CP violation. From the moments of the Kin distribution one obtains $\sin(2\beta) = 0.604 \pm 0.088$. A likelihood analysis of the same sample yields $\sin(2\beta) = 0.590 \pm 0.086$. The slight decrease of the statistical uncertainty achieved by using a likelihood analysis with this particular sample of events is itself a statistical fluctuation: on average, a likelihood analysis uncertainty would lead to a reduction of the statistical uncertainty by a factor 0.995.

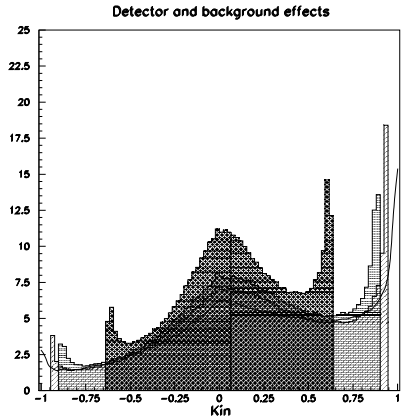

**Figure 2.** *Examples of $\mathcal{K}$ distributions in presence of experimental limitations. The curve is the $\Psi(\mathcal{K})$ corresponding to the ideal situation of Fig.1. When experimental effects are taken into account the wings of the distributions no longer reach the extremal values $\mathcal{K} = \pm 1$. The effects of $\sigma_z$, $B_S$ and $w_{\text{tag}}$ are added one after the other (see text).*

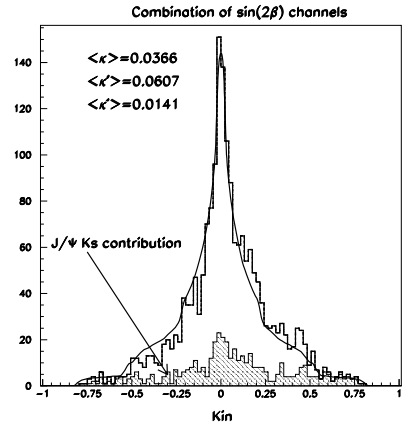

**Figure 3.** *Monte Carlo simulation of 2000 events corresponding to the with the input value $\sin(2\beta) = 0.7$. From the moments of the Kin distribution, one obtains $\sin(2\beta) = 0.604 \pm 0.088$. A likelihood analysis of the same sample yields $\sin(2\beta) = 0.590 \pm 0.086$. The curve represents the $\Psi_0(\mathcal{K})$ function, the moments of which are $\langle \mathcal{K} \rangle = 0$, $\langle \mathcal{K}^2 \rangle = 0.621$, and $\langle \mathcal{K}^4 \rangle = 0.0150$.*

Because the properties of the Kin variable follows from the identity Eq. (5), which is valid in any case, all channels corresponding to $\sin(2\beta)$ can be merged together into a single distribution. For modest statistics, while the event samples are not sufficient to perform meaningful channel by channel (or experiment

## 4. Conclusion

Although the measurement of $\sin(2\beta)$ can be extracted from a likelihood analysis using the time distribution of events, the variable which contains all the information on CP violation is not the time, but the Kin variable. This variable provides a tool to graphically display CP violation, to identify the most potent events, and to merge together a variety of channels and experiments. Furthermore, from the moments of its distribution, one may probe CP violation in a straightforward way and determine $\sin(2\beta)$ with an accuracy equivalent to the one of a likelihood analysis. This optimal variable appears particularly adequate for the forthcoming experimental analyses, aiming at establishing CP violation in $B$ decays.